# A 1 GHZ RF TRIGGER UNIT IMPLEMENTED IN FPGA LOGIC*

D. Barrientos†, J. Molendijk, G. Hagmann, CERN, Geneva, Switzerland

*Abstract*

Applications of Trigger Units (TU) can be found in almost all accelerators at CERN. The requirements in terms of operating frequencies, configuration or modes of operation change from one application to another, however, in terms of design requirements for the Trigger Unit, the operating frequency is probably the most demanding one. In this work, we present an implementation of a Trigger Unit almost fully embedded in the FPGA logic operating at a maximum frequency of 1 GHz using the internal serializer/deserializer circuitry to simplify the timing constraints of the design. This implementation allows easy reconfiguration of the module and the development of new modes of operation, which are described in this paper.

## INTRODUCTION

In LLRF systems at CERN, Trigger Units (TU) generate pulses synchronized with a RF signal. They can generate single pulses to be used in beam observation systems, trains of pulses separated by a certain number of RF periods or infinite train of pulses to generate, for example, the revolution frequency signal of a circular accelerator. Applications of these modules are found in several subsystems in the CERN accelerator complex for timing, observation and synchronization purposes.

An example of the use for a TU is depicted in Figure 1. The *Sync* pulse start a counter with the Bunch number (*B*), which selects a specific bunch in the circular accelerator. Then, a counter with the Harmonic number (*H*) of the accelerator multiplied by the selected Number of turns (*T*) allows to produce an *Out* pulse at the selected bunch, *T* turns after the *Sync* pulse. Another example, generating four output pulses is presented in a chronogram in Figure 2.

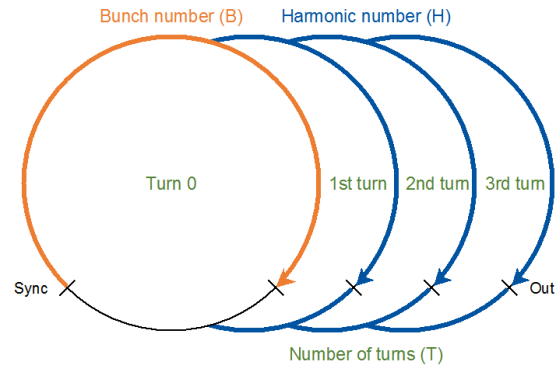

Figure 1: Principle of operation of a Trigger Unit.

TU units are present in LLRF systems at CERN since the end of the 1970s. The last module developed for this purpose was the VME Trigger Unit (VTU) [1], which was designed for the LLRF system of the Large Hadron Collider (LHC), more than 10 years ago. Nowadays, the VTU is being used in several systems throughout the CERN accelerator complex.

However, the production and maintenance of VTU modules is becoming difficult due to the obsolescence of some components.

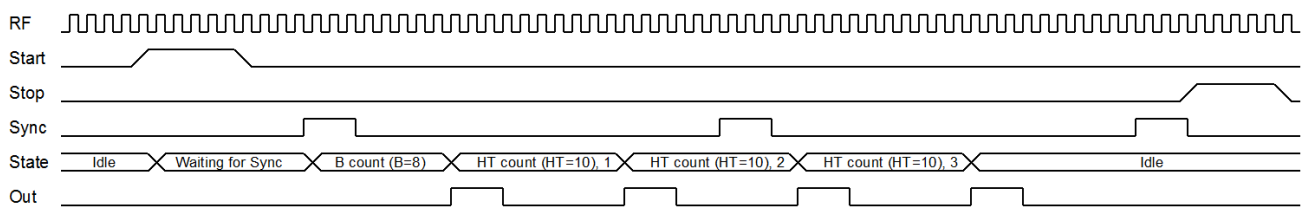

Figure 2: Chronogram with Trigger Unit signals. Windowed operation mode: B=8, HT=10, W=4.

## HARDWARE PLATFORM

For the implementation of a new Trigger Unit, we have reused an existing module, the hardware of which suits the TU requirements, as shown in Figure 3. The Chopper Trigger Unit (CTU) module [2-3] has been designed to control the beam chopper located after the RFQ section in the Linac4 accelerator, currently being commissioned at CERN. The module is used to generate chopping patterns synchronously to the Linac4 RF frequency (352.2 MHz) and the PS-Booster revolution frequency, allowing bunch to bucket transfer during injection.

As the hardware features of the CTU module are compatible with its application as a TU, its FPGA can be reconfigured for this purpose. The *Sync* delay line (see Figure 3) can be used to reset the external clock divider to have a deterministic relation between *Sync* and *RF* inputs to avoid metastability issues. The *Out* delay can be used for fine synchronization of the output signal.

___________________________________________

* The authors of this work grant the arXiv.org and LLRF Workshop's International Organizing Committee a non-exclusive and irrevocable license to distribute the article, and certify that they have the right to grant this license.

† diego.barrientos@cern.ch

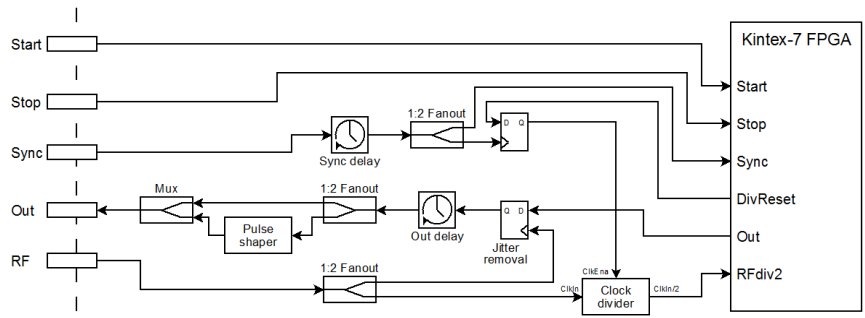

Figure 3: Simplified block diagram of CTU hardware.

## FPGA IMPLEMENTATION

If a TU needs to work with RF frequencies up to about 200/300 MHz, modern FPGAs would be able to implement the counters and state machine logic needed to produce pulses synchronously to the RF (clock) signal after a configurable number of clock cycles. However, when trying to deal with RF frequencies up to 1 GHz, as the VTU module can handle, this approach cannot meet internal FPGA timing requirements and, therefore, a different methodology is needed.

The implementation of a TU in the CTU module takes advantage of the embedded serializer and deserializer circuitry available near the input-outputs of the Xilinx Kintex-7 FPGA, as shown in Figure 4. The *RF* signal is externally divided by 2 and enters the FPGA with a maximum frequency of 500 MHz. This signal is used to deserialize the *Sync* input at Double Data Rate (DDR), using both edges of the divided *RF* signal and, therefore, sampling the *Sync* signal at a maximum rate of 1 Gbps. The output of the deserializer is an 8-bit parallel bus clocked with a divided copy of the *RF* signal with a maximum frequency of 125 MHz.

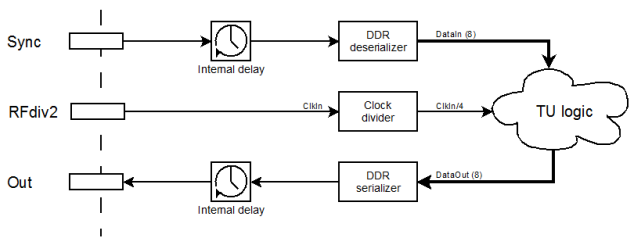

Figure 4: Block diagram of internal signal paths.

A similar mechanism is used for the output stage, where an 8-bit parallel bus is serialized to produce an output stream at a maximum rate of 1 Gbps. This strategy relaxes the timing constraints of the internal logic in the FPGA, but complicates the design of the TU logic block that transforms the input to output data buses.

To validate this methodology, specific tests have been performed to guarantee that the delay between the *Sync* and *Out* signals at the interface of the module remains constant when the same configuration is applied. To achieve this result, the reset scheme of clock dividers (either internal or external to the FPGA), serializer and deserializer blocks has been carefully designed. Validation tests included several power cycles of the module without observing delay variations.

In addition, the peak-to-peak jitter of the *Out* signal with respect to the *RF* signal has been measured for both VTU and CTU modules, obtaining a value of 96.1 ps and 39.3 ps respectively. The standard deviation of the jitter for the CTU module is 4.4 ps.

## MODES OF OPERATION

The new implementation of a TU in the CTU module has 6 modes of operation. All of them are governed by slow *Start* and *Stop* triggers that prepare the TU to accept the next Sync pulse or to end the current execution respectively.

1. *Single pulse*: This mode is used to generate only one pulse at the output, *B* RF periods after the *Sync* pulse.
2. *Infinite window*: In this mode, the first pulse is also generated *B* RF periods after the *Sync* pulse but the next pulses are separated by *HT* cycles until it is stopped.
3. *Windowed operation*: This mode is similar to *Infinite window*, but having the number of output pulses limited to the *W* value.

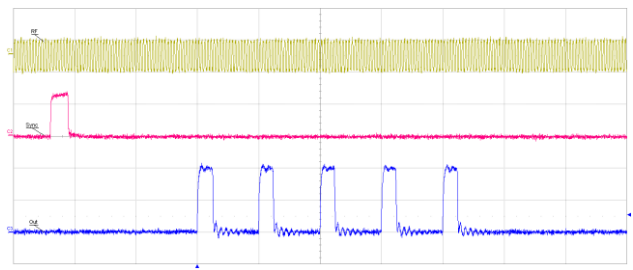

Figure 5: *Windowed operation* mode: B=8, HT=20, W=5. Timebase: 20 ns/div. *RF* (yellow) frequency: 1 GHz. Difference between two *Out* pulses (blue): 20 ns (20 *RF* periods).

4. *SyncLess operation*: Similarly to the *Infinite window* mode, an infinite train of pulses separated by *HT* RF periods is generated. However, the *Sync* pulse is not required and the *Start* signal triggers the production of output pulses. A typical use of this mode is the generation of the master revolution frequency train.

5. *Low frequency generation*: The output signal of this mode is not a pulse but a square wave starting *B* RF cycles after the *Sync* pulse. Then, the *HT* value is used to indicate the number of RF periods of the high and low values. Therefore the output signal has a frequency of $f_{RF}/(2\ HT)$. A configurable flag can be used to generate unbalanced square signals with one cycle more in the low state produces a wave with a frequency of $f_{RF}/(2\ HT+1)$.

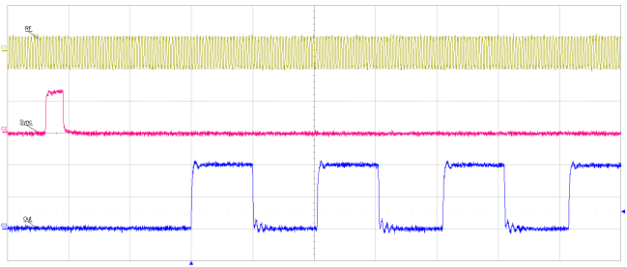

Figure 6: *Low frequency generation* mode: B=8, HT=20 (switching). Timebase: 20 ns/div. *RF* (yellow) frequency: 1 GHz. Frequency of *Out* signal (blue): 24.39 MHz (41 *RF* periods).

6. *Play memory*: In this mode, the user can fill an 8 kB memory to produce a pattern synchronous to the RF signal at the output. The playback length is programmable up to the bit level and determines the pattern repetition rate. This mode may be used to generate a simulated beam pattern of a circular accelerator.

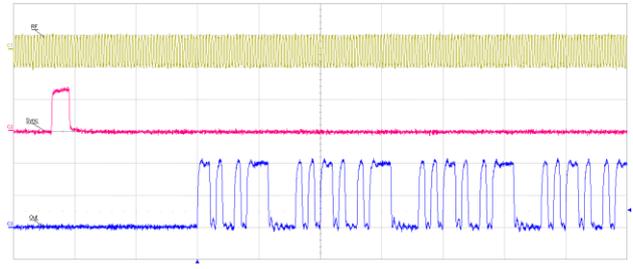

Figure 7: *Play memory* mode: B=8, memory contents. Timebase: 20 ns/div. *RF* (yellow) frequency: 1 GHz. *Out* (blue) pattern length: 40 bits (repeating every 40 ns).

## CONCLUSION

A new implementation of a Trigger Unit has been presented making use of the embedded serializer and deserializer circuits of the Xilinx Kintex-7 FPGA. This technique has been proven to cover all TU requirements while simplifying the design and allowing to implement new modes of operation. The implementation of the TU in the CTU module is planned to be commissioned in different accelerators of the CERN complex during the following months.